\documentclass[aps,prl,twocolumn,amsmath,amssymb,nofootinbib,superscriptaddress]{revtex4-1}

\usepackage{times}
\usepackage[pdftex]{graphicx}
\usepackage{dcolumn}
\usepackage{bm}
\usepackage{amsmath}
\usepackage{indentfirst}
\usepackage{float}
\usepackage[dvipsnames]{xcolor}

\begin{document}

\title{A loophole-free Wheeler-delayed-choice experiment}

\author{H.-L. Huang}

\affiliation{Hefei National Laboratory for Physical Sciences at Microscale and Department of Modern Physics,\\
University of Science and Technology of China, Hefei, Anhui 230026, China}
\affiliation{CAS Centre for Excellence and Synergetic Innovation Centre in Quantum Information and Quantum Physics,\\
University of Science and Technology of China, Hefei, Anhui 230026, China}
\affiliation{Henan Key Laboratory of Quantum Information and Cryptography, Zhengzhou, Henan 450000, China}

\author{Y.-H. Luo}

\affiliation{Hefei National Laboratory for Physical Sciences at Microscale and Department of Modern Physics,\\
University of Science and Technology of China, Hefei, Anhui 230026, China}
\affiliation{CAS Centre for Excellence and Synergetic Innovation Centre in Quantum Information and Quantum Physics,\\
University of Science and Technology of China, Hefei, Anhui 230026, China}

\author{B. Bai}
\author{Y.-H. Deng}
\author{H. Wang}
\author{H.-S. Zhong}
\author{Y.-Q. Nie}
\author{W.-H. Jiang}
\author{X.-L. Wang}
\author{J. Zhang}
\author{Li Li}
\author{Nai-Le Liu}
\affiliation{Hefei National Laboratory for Physical Sciences at Microscale and Department of Modern Physics,\\
University of Science and Technology of China, Hefei, Anhui 230026, China}
\affiliation{CAS Centre for Excellence and Synergetic Innovation Centre in Quantum Information and Quantum Physics,\\
University of Science and Technology of China, Hefei, Anhui 230026, China}

\author{Tim Byrnes}
\affiliation{New York University Shanghai, 1555 Century Ave, Pudong, Shanghai 200122, China}
\affiliation{NYU-ECNU Institute of Physics at NYU Shanghai, 3663 Zhongshan Road North, Shanghai 200062, China}
\affiliation{Department of Physics, New York University, New York, NY 10003, USA}

\author{J. P. Dowling}
\affiliation{Hearne Institute for Theoretical Physics, Department of Physics and Astronomy, Louisiana State University, Baton Rouge, Louisiana 70803, USA}
\affiliation{Hefei National Laboratory for Physical Sciences at Microscale and Department of Modern Physics,\\
University of Science and Technology of China, Hefei, Anhui 230026, China}
\affiliation{NYU-ECNU Institute of Physics at NYU Shanghai, 3663 Zhongshan Road North, Shanghai 200062, China}

\author{Chao-Yang Lu}
\affiliation{Hefei National Laboratory for Physical Sciences at Microscale and Department of Modern Physics,\\
University of Science and Technology of China, Hefei, Anhui 230026, China}
\affiliation{CAS Centre for Excellence and Synergetic Innovation Centre in Quantum Information and Quantum Physics,\\
University of Science and Technology of China, Hefei, Anhui 230026, China}

\author{Jian-Wei Pan}
\affiliation{Hefei National Laboratory for Physical Sciences at Microscale and Department of Modern Physics,\\
University of Science and Technology of China, Hefei, Anhui 230026, China}
\affiliation{CAS Centre for Excellence and Synergetic Innovation Centre in Quantum Information and Quantum Physics,\\
University of Science and Technology of China, Hefei, Anhui 230026, China}

\date{\today}


\begin{abstract}
Wheeler's delayed-choice experiment investigates the indeterminacy of wave-particle duality and the role played by the measurement apparatus in quantum theory. Due to the inconsistency with classical physics, it has been generally believed that it is not possible to reproduce the delayed-choice experiment using a hidden variable theory.  Recently, it was shown that this assumption was incorrect, and in fact Wheeler's delayed-choice experiment can be explained by a causal two dimensional hidden-variable theory [R. Chaves, G. B. Lemos, and J. Pienaar, Phys. Rev. Lett. 120, 190401 (2018)]. Here, we carry out an experiment of a device-independent delayed-choice experiment using photon states that are space-like separated, and demonstrate a loophole-free version of the delayed-choice protocol that is consistent with quantum theory but inconsistent with any causal two-dimensional hidden variable theory. This salvages Wheeler's thought experiment and shows that causality can be used to test quantum theory in a complementary way to the Bell and Leggett-Garg tests.
\end{abstract}

\maketitle

After the two famous Bohr-Einstein debates of 1927 and 1930 on the validity of the quantum theory \cite{bohr1996discussion}, Einstein had to accept that quantum mechanics was correct. However, in his paper with Podolsky and Rosen (EPR) \cite{Einstein1935}, EPR claimed that quantum theory, while not incorrect, was incomplete. That paper showed that quantum-entangled states had elements of nonlocality, un-reality, and uncertainty that no ``sensible'' theory should have. The EPR paper opened the door for a replacement theory for quantum mechanics, now called a hidden variable (HV) theory, that would be more like a classical statistical theory, where the statistics were governed by HVs that were either unknown or unaccessable. Von Neumann then provided a proof that no HV theory could reproduce all the predictions of quantum theory \cite{neumann1932}. Later, Bohm produced a HV theory that reproduced all the predictions of nonrelativistic quantum theory. To solve this apparent paradox, Bell showed that von Neumann had made an explicit assumption that the HV theory was local, but that the HV theory of Bohm was nonlocal --- that is actions at one place could affect outcomes far away in apparent violation of Einstein causality \cite{Bell1965}. Thus, it is important to make the locality requirement explicit and show whether local HV theory could reproduce the predictions of quantum theory.

\begin{figure}[tb]
\includegraphics[width=\columnwidth]{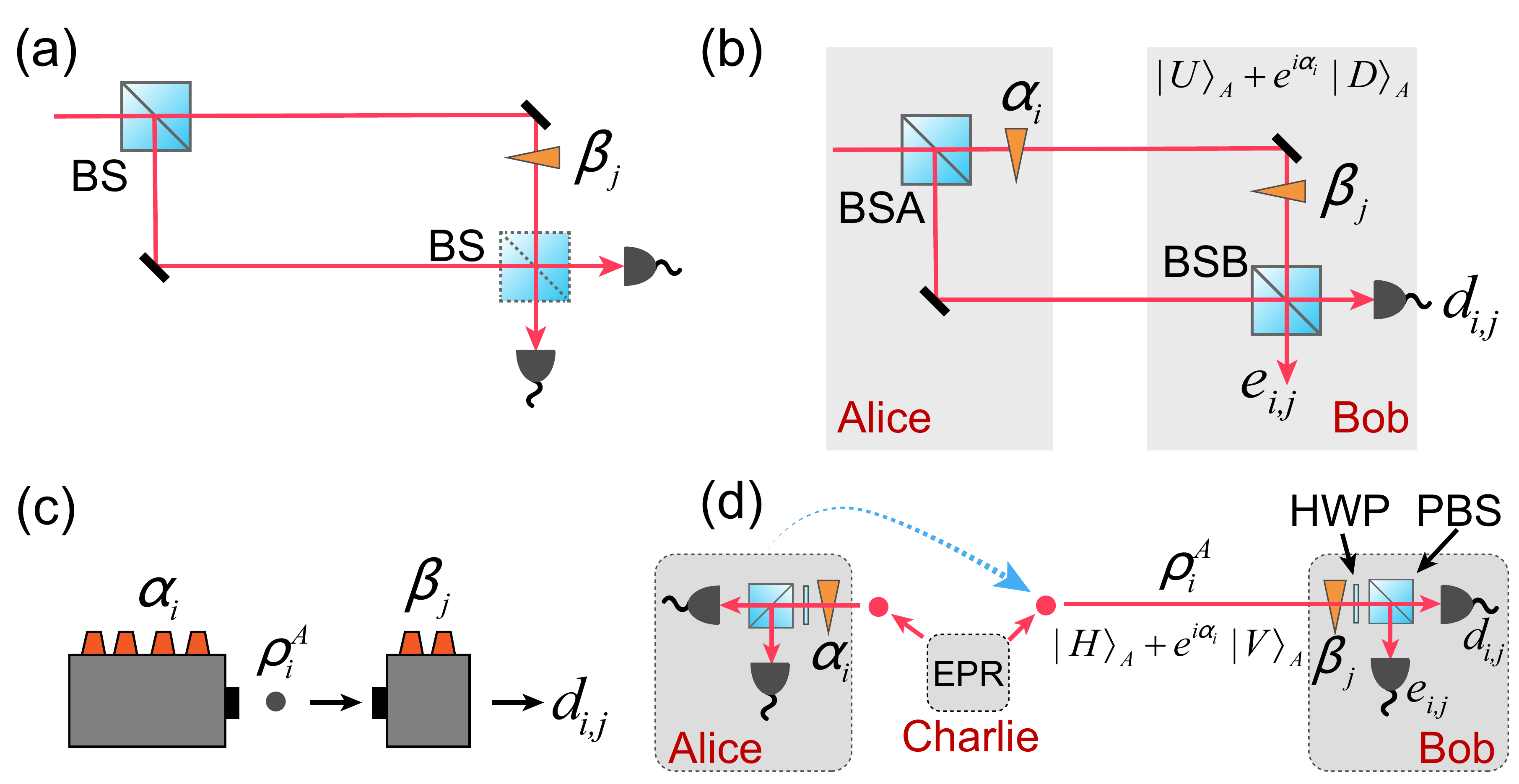}
\caption{Various configurations of Wheeler's delayed choice experiment (WDCE).  
(a) The original WDCE implemented with photons. Wave-like or particle-like behavior can be produced by either putting in or leaving out the lower beamsplitter (BS).  (b) The modified WDCE (MWDCE). The gray areas correspond to Alice (Preparer) and Bob (Measurer), respectively.  In the modified setup, the second beamsplitter is always put in place, Alice has the ability to prepare a set of states, controlled by her own phase shifter ${\alpha_i} \in \{ {\alpha_0} = 0,{\alpha_1} = \pi ,{\alpha_2} =  - \pi /2,{\alpha_3} = \pi /2\}$, and two types of detection is controlled by the phase shifter ${\beta _j} \in \{ {\beta _0} = \pi /2,{\beta _1} = 0\}$ of Bob. The $U$ is the upper path and $D$ is the lower path.  (c) The device-independent prepare-and-measure (PAM) scenario for the MWDCE. An initial black-box prepares different physical systems (upon pressing a button labeled by ${\alpha _i}$) that are then sent to a second black-box where the systems are measured (upon pressing a button labeled by ${\beta _j}$) to produce an outcome labeled by $d_{ij}$. (d) Schematic experimental description performed in this paper. Alice prepares a photon by measuring an EPR state in one of two basis, producing four possible states $ | H \rangle + e^{i \alpha_i} | V \rangle $.  Bob then applies a phase shift and interferes the photon using a polarizing beam splitter (PBS) to obtain the output.} \label{fig:1}
\end{figure}

In an effort to challenge quantum mechanics, and make quantum predictions consistent with common sense, it was suggested that quantum particles can actually know in advance to which experiment they are going to be confronted through a HV, and thus can determine which behavior to show. For example, the photon could ``decide'' whether it was going to behave as a particle or behave as a wave before it reach the detection device in the double-slit experiment. However, Wheeler published two theory papers, now called  Wheeler's delayed-choice experiments (WDCE), that claimed to exclude the causal link between the experimental setup and a HV that predefines the photons behavior, and point out that complementarity and wave-particle duality alone contained an element of Einstein nonlocality \cite{Wheeler1978, Wheeler1984}. Fig. \ref{fig:1}(a) shows an example of an implementation of WDCE, where a photon enters an Mach-Zehnder interferometer.  Wave-like behavior or particle-like behavior can be dictated according to whether the second beamsplitter (controlled by Bob) is put in place or not.  
So far, WDCE has been implemented experimentally in various quantum systems \cite{hellmuth1987delayed, lawson1996delayed, kim2000delayed, zeilinger2005happy, jacques2007experimental, Vedovatoe1701180}. Interestingly, a recent extension, quantum delayed-choice experiment (QDCE), suggested using a quantum beam splitter at the interferometer's output \cite{ionicioiu2011proposal, ionicioiu2014wave}, enabling us to project the test photon into an arbitrary coherent wave-particle superposition, which motivated many QDCE experiments \cite{kaiser2012entanglement, peruzzo2012quantum, rab2017entanglement, Liue1603159}. In short, until recently experimental demonstrations of WDCE (or QDCE) were thought to have perfectly ruled out the possible of quantum behavior induced by HV.

Chaves, Lemos, and Pienaar recently revisited WDCE, and showed using methods in causal inference \cite{pearl2009causality, Ringbauere1600162} that the original WDCE and QDCE can in fact be modeled by a causal HV theory \cite{Chaves2018}. The HV theory they suggest follows the same causal structure of the experiment shown in Fig. \ref{fig:1}(a), such that statistics produced by the final detection can be determined from a HV and the information of the type of measurement being performed.  To overcome this shortcoming, the authors suggest a modified version of WDCE (MWDCE, Fig. \ref{fig:1}(b)) that cannot be explained by a causally structured HV theory, assuming a HV dimension of two. In the modified setup, the second beamsplitter is always put in place, and the two types of detection is controlled by the phase shifter $ \beta_j $.  The additional component is that Alice has the ability to prepare a set of states, controlled by her own phase shifter $ \alpha_i $.  Here, an important aspect is that Alice's setting $ \alpha_i $ cannot be directly communicated to Bob.  In Ref. \cite{Chaves2018} it is shown that this new setup can no longer be modeled by a causally structured HV theory, as long as the HV has a dimension of two. Such a causal HV theory can be distinguished from a genuine quantum theory by comparing the statistics to a device-independent prepare and measure (PAM) scenario \cite{Bowles2014, gallego2010device} (Fig. \ref{fig:1}(c)).  Using the statistics of the measurements and the setting values, a device-independent witness \cite{Bowles2014} can be constructed that is capable of distinguishing between a causally structured HV theory. The dimensional witness can, in principle, work with any transmittance strictly larger than zero, thus making MWDCE specially suited to loophole-free experimental implementations \cite{giustina2015significant, shalm2015strong}.

In this paper, we carry out a device-independent demonstration of the MWDCE. 
By measuring a device-independent witness we verify that two-dimensional causally structured HV theory can indeed be ruled out.  We furthermore rule out any a priori prepared correlated noise sources that may exist between Alice and Bob by violation of a dimension witness inequality.  Finally, we quantify the degree of retrocausality that would be required to reproduce our experiments, placing bounds on such hypothetical scenarios. Our experiment tests the causal structure in a WDCE, and is complementary to tests of nonlocality in quantum theory such as in a Bell test or a Leggett-Garg test \cite{Leggett1985}. 

\begin{figure*}
\includegraphics[width=1.5\columnwidth]{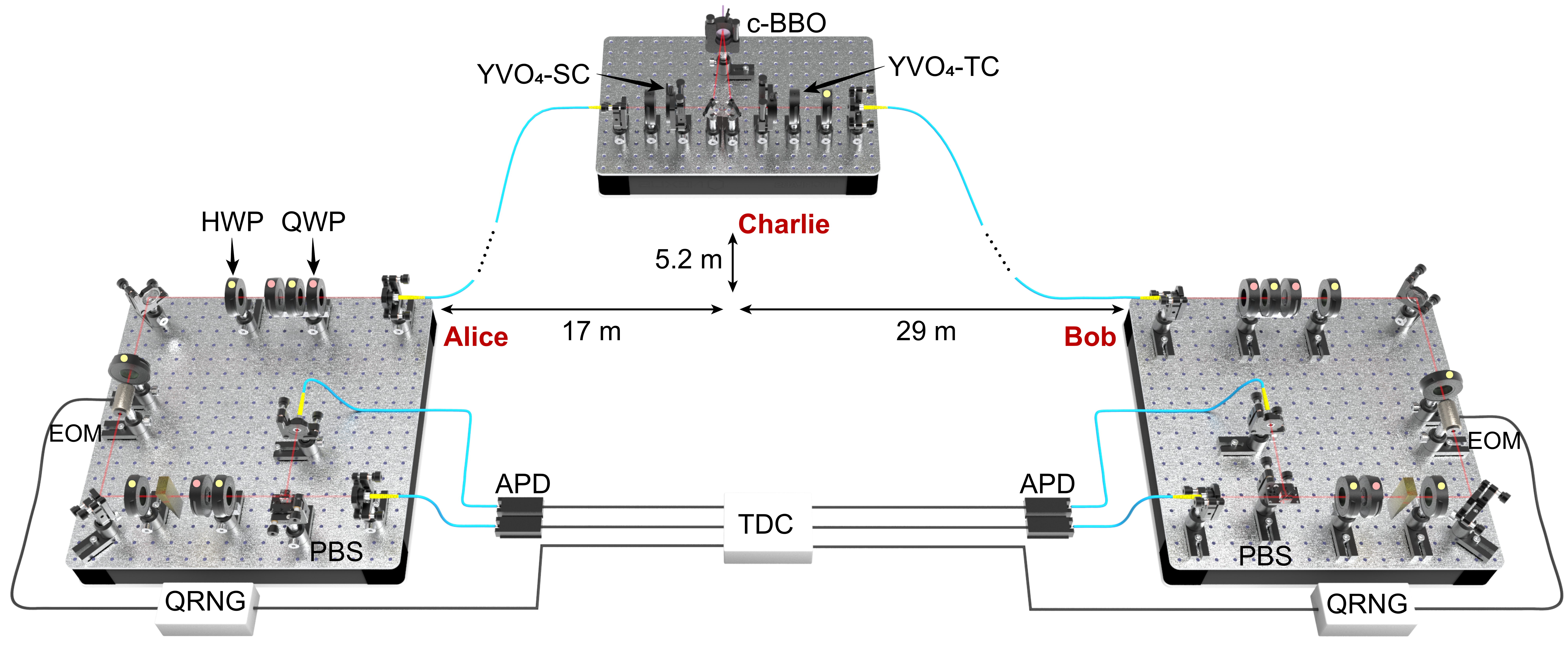}
\caption{Experimental setup. The polarization-entangled photon pairs produced by SPDC are coupled into the single-mode fiber and sent to Alice (preparer) and to Bob (measurer), respectively. The distance between Alice and Bob is 46 meters and the length of the fiber from the entanglement source (Charlie) to Alice is shorter than that to Bob (the fiber lengths of Charlie-Alice and Charlie-Bob are 28 meters and 33 meters, respectively). Alice's photon passes through an EOM, then is measured by the APD which collpases Bob's state to $|H{\rangle _B} + {e^{i{\alpha _j}}}|V{\rangle _B}$ with $ \alpha_j \in \{0,\pi,\pm \pi/2 \} $, thereby preparing the state. Bob's photon then passes through another EOM where a random phase $ \beta_j \in \{ \pi/2, 0 \} $ is applied and is interfered with a PBS, followed by a measurement by the APD. The measurement basis of Alice and Bob are each randomly determined by two independent and space-like separated QRNGs and EOMs. In order to meet the delayed-choice condition, the measurement basis of Bob is chosen much later than that Alice, so that no causal information can reach Bob before his QNRG has fired.} \label{fig:2}
\end{figure*}


\begin{figure}[t]
\includegraphics[width=\columnwidth]{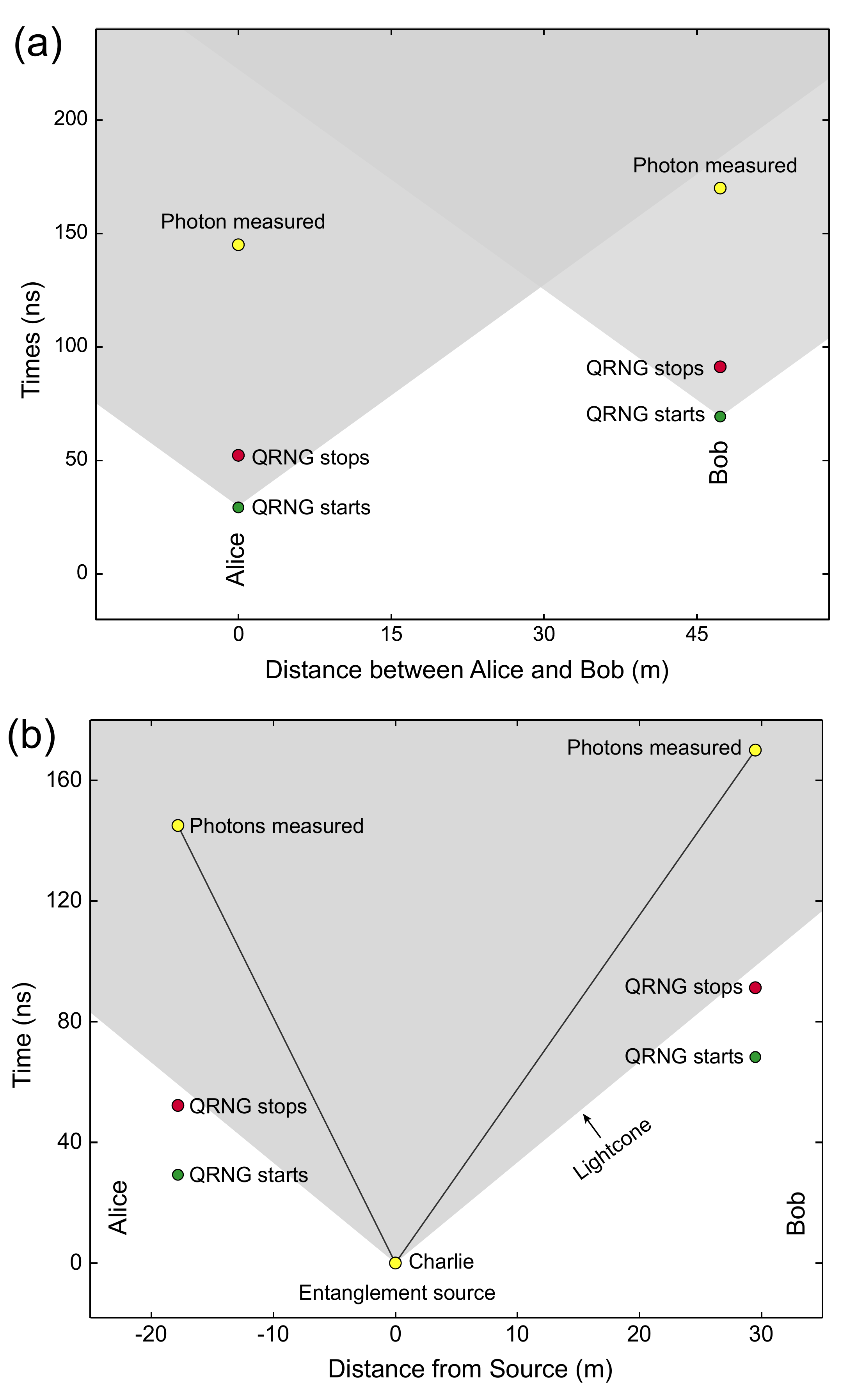}
\caption{Minkowski diagrams for the spacetime events related to Alice, Bob, and the source Charlie. All light cones are shaded gray.  (a) Alice and Bob are space-like separated as the measurement is finished by Alice and Bob before information about the other party's measurement setting could have arrived. (b) The quantum random number generators (QRNGs) at Alice and Bob finish picking a setting outside the light cone of the generation of an entangled photon pair by Charlie. All the events in our experiment are space-like separated.} \label{fig:3}
\end{figure}

Figure \ref{fig:1}(d) shows the schematic experimental implementation for realizing the MWDCE.  The path-based interferometer as shown in Fig. \ref{fig:1}(b) is implemented in our experiment by a polarization-based interferometer.  The horizontal ($H$) and vertical ($V$) polarizations correspond to the upper and lower paths in the MZIs in Fig.~\ref{fig:1}(b). To prepare the various initial states we use an EPR photon pair emitter located at Charlie's location $C$. Charlie is located closer to Alice than to Bob to allow Alice to make her preparation step first by measuring her half of the entangled pair after applying her polarization rotation ${\alpha _i}$ first. This collapses Bob's half of the entangled pair into a single photon to which Alice's polarization rotation ${\alpha _i}$ has been applied. Then Bob applies his polarization rotation ${\beta _j}$, before any influence from Alice can reach him, and performs the interference by passing the photon through a polarizing beam splitter (PBS), followed by a measurement.  We note that this is much different than the usual Bell test where Alice and Bob are symmetric in the protocol and measure correlations between their outcomes as opposed to the current scenario, and Bob measures an outcome condidtioned on Alice's preparation state. Time ordering is now important in the current scheme.

Our experimental setup is shown in Fig. \ref{fig:2}. At Charlie's location, an ultraviolet laser pulse with a central wavelength of 394 nm, pulse duration of 150 fs, and repetition rate of 80 MHz passes through a $\beta$-barium borate (BBO) crystal to produce a polarization-entangled pairs $|{\Psi ^ + }{\rangle _{AB}} = |H{\rangle _A}|V{\rangle _B} + |V{\rangle _A}|H{\rangle _B}$ \cite{wang2016experimental}. A half-wave plate (HWP) is placed at an arm of the entangled pairs to produce the Bell state  $|{\Phi ^ + }{\rangle _{AB}} = |H{\rangle _A}|H{\rangle _B} + |V{\rangle _A}|V{\rangle _B}$. 
The two photons are then coupled to the into single mode fiber and sent to Alice (preparer) and Bob (measurer), respectively. The distance between Alice and Bob is 46 m, and the length of the fiber from the entanglement source (Charlie) to Alice is shorter than that to Bob. The electro-optic modulator (EOM) then applies a phase shift of 0  or $ \pi/2 $ chosen by a quantum random number generator (QRNG) such that Alice's photon is measured as one of four possibilities  $ |H \rangle_A \pm |V \rangle_A $ and $ |H \rangle_A \pm i |V \rangle_A $.  This causes Bob's state to collapse to $ |H \rangle_B \pm |V \rangle_B $ or $ |H \rangle_B \mp i |V \rangle_B $ respectively.  Bob then deploys his delayed-choice of applying two polarization rotations ${\beta _j} \in \{ {\beta _0} = \pi /2,{\beta _1} = 0\}$, chosen by another QRNG.  This is equivalent to measuring his photon one of the two bases $\{ |H{\rangle _B} \pm {e^{i{\beta _j}}}|V{\rangle _B} \}$ chosen by $\beta_j $.  We note that the four preparation states are produced by a combination of Alice's QRNG and the random collapse of the 
EPR state.  Thus no postselection is performed to prepare the state.


In order that there is no causal connection between Alice's choice of settings and Bob's detection, they are space-like separated at large enough distances such that the QRNG and the photon measurements are both outside their mutual lightcone (Fig. \ref{fig:3}(a)).  
In other words, in order to meet the delayed-choice condition, the measurement basis of Bob must be chosen sufficiently later than the measurement basis choice of Alice. Two synchronized signals from a central clock are used to trigger the two QNRGs. In addition, the QNRGs must fire outside the light cone of the EPR source to ensure that Charlie cannot influence the outcome of the QNRGs, as shown in Fig.~\ref{fig:3}(b). Without this additional condition on the EPR source, the experiment does not meet all the assumptions of the PAM scenario.

\begin{figure}[t]
\includegraphics[width=\columnwidth]{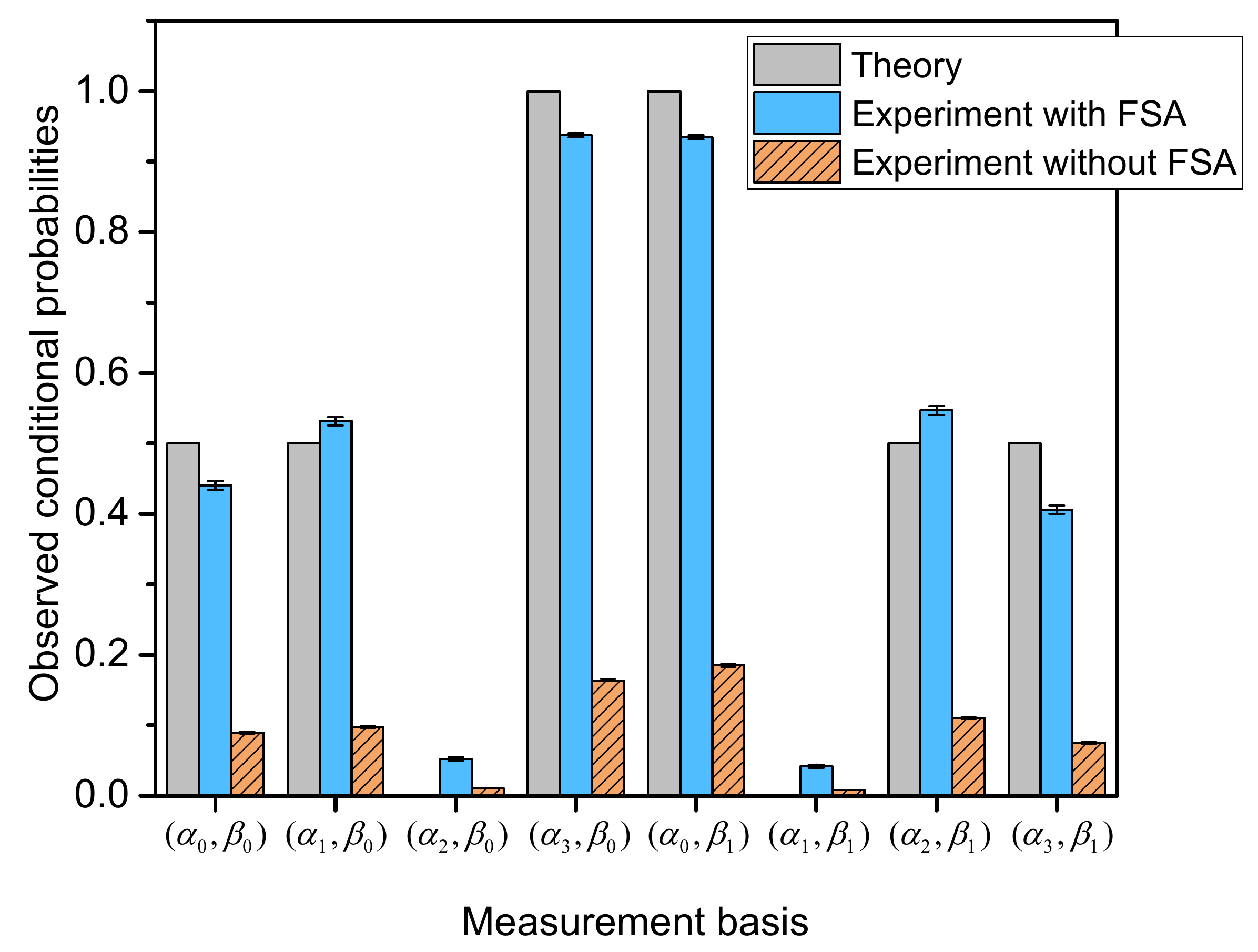}
\caption{Conditional probabilities for the outcome $ |H{\rangle _B} \pm {e^{i{\beta_j}}}|V{\rangle _B} $ by Bob for various 
settings $ (\alpha_i, \beta_j ) $.  Error bars represent one standard deviation, deduced from propagated Poissonian counting statistics of the raw detection events.} \label{fig:4}
\end{figure}

To measure the causal HV witness, we measure the conditional probability of the detection of the state $ |H{\rangle _B} \pm {e^{i{\beta_j}}}|V{\rangle _B} $ as a function of the settings $ ( \alpha_i, \beta_j ) $.  The matrix elements of the $(2 \times 2)$ witness matrix $W$ is defined as \cite{Bowles2014} 
\begin{align}
W_{k,l}= p(d_{2k-2,l-1}) - p(d_{2k-1,l-1}),
\end{align}
where $k,l \in \{ 1,2\} $, $ p(d_{i,j}) $ is the probability of the outcome $ d_{ij} $.  For a causally structured two dimensional HV theory one should find that $ \det (W) = 0 $, whereas according to quantum mechanics for an ideal system $ |\det (W)| = 1 $. Figure \ref{fig:4} shows a comparison of the theoretical predictions and  the experimental measurements with and without the fair-sampling assumption (FSA), which performs a postselection only on coincidence events. With the FSA, the two-dimensional witness is calculated as $|\text{det}({W})| = 0.778 \pm 0.005$ according to the measurement results. Experimental errors mainly come from higher-order events in the SPDC and the control accuracy of EOM.  Without the FSA, the two-dimensional witness is measured to be $|\text{det}({W})| = 0.0268 \pm 0.0006$, which is mainly due to the low collection efficiency. However, even without the FSA, the witness of the two dimensional HV,  $|\text{det}({W})| = 0$, is still violated by 44 standard deviations.   This means that we have --- with a great degree of confidence --- shown that our experiment is inconsistent with a causally structured HV theory with two dimensions. Thus, this scheme is highly resilient to detection inefficiencies.



\begin{figure}[t]
\includegraphics[width=\columnwidth]{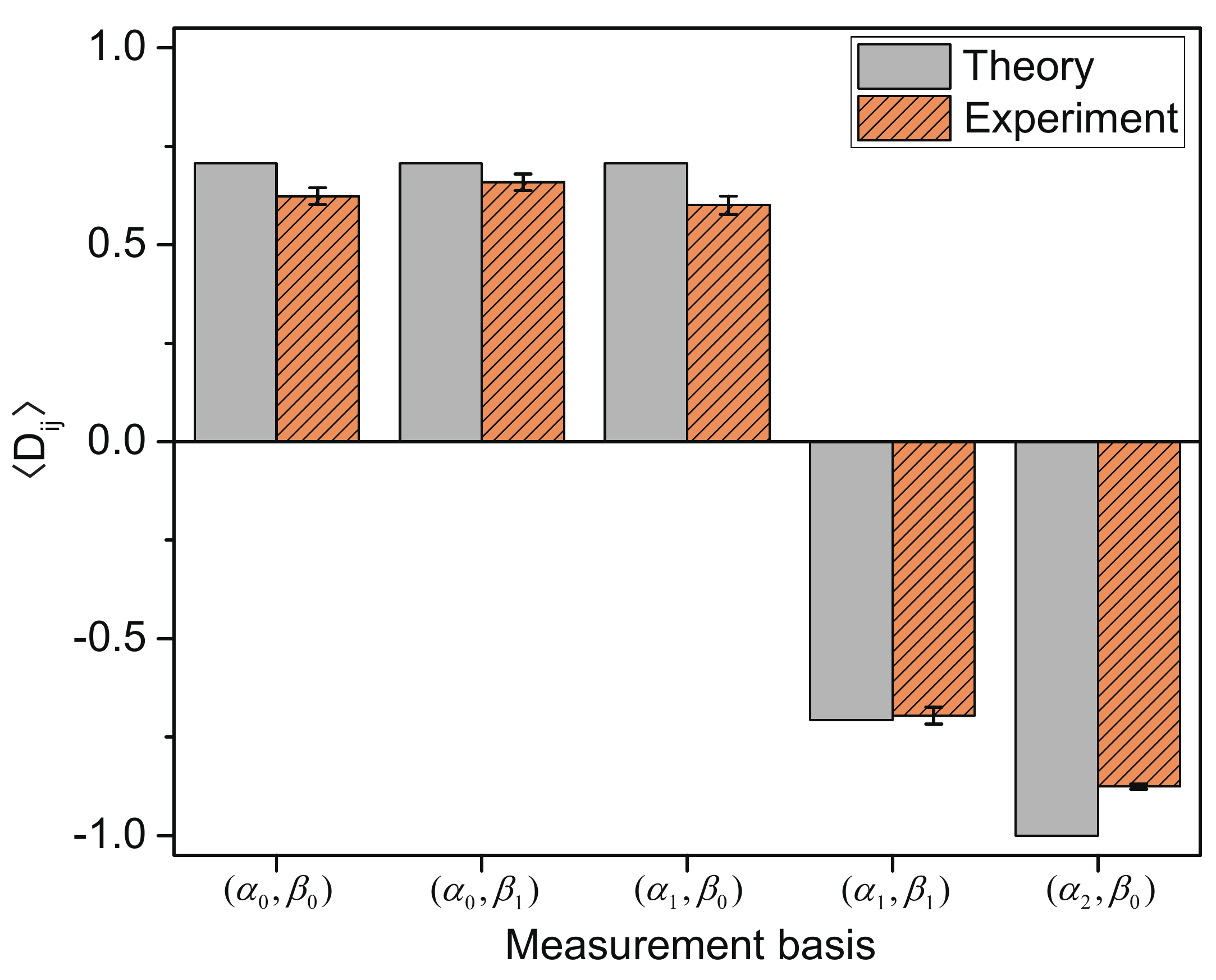}
\caption{Experimental results of the dimensional witness $I_{DW}$. The $\langle {D_{ij}}\rangle$ for the theoretical predictions and experimental measurement are shown as grey and yellow bars, respectively. Error bars represent one standard deviation, deduced from propagated Poissonian counting statistics of the raw detection events.} \label{fig:5}
\end{figure}

We also test for HV models that are dependent upon hidden noise terms that could influence the output of the interferometer.   While our experiment rules out the causal influence of the operations that Alice and Bob perform on each other, it is possible that such a noise variable could be prepared long before the start of the experiment, and would not be forbidden by causality.  We use the dimension witness  \cite{gallego2010device,ahrens2012experimental}
\begin{align}
{I_{DW}} = \langle {D_{00}}\rangle  + \langle {D_{01}}\rangle  + \langle {D_{10}}\rangle  - \langle {D_{11}}\rangle  - \langle {D_{20}}\rangle
\end{align}
where  $ \langle {D_{ij}}\rangle  = p(e_{i,j}) - p(d_{i,j}) $, $ p(e_{i,j}) $ is the probability of the outcome $ e_{ij} $ (see Fig. \ref{fig:1}(d)), and the measurement settings are  $ \alpha_i \in  \{ \pi/4,  3\pi /4, - \pi /2 \} $ and  $\beta_j \in \{\pi /2,  0 \}$ as before.   Any HV theory that accounts for correlations between Alice and Bob hiding in the noise gives a strict bound of $I_{DW}^{\text{HV}} \le 3$, with quantum theory predicting $I_{DW}^Q = 1 + 2\sqrt 2  \approx 3.828$ \cite{Chaves2018, tavakoli2018self}. The data for this witness is shown in Fig. \ref{fig:5}, and yields, $I_{DW}^\text{Q} = 3.445 \pm 0.043$, with the FSA. The bound in this case is violated by 10 standard deviations.


The above shows that a causally structured two-dimensional HV model would be inconsistent with our experimental results.  However, if one allows for the possibility of retrocausality (i.e. signaling backwards in time), it becomes possible to construct a HV theory that can account for statistics consistent with quantum experiment.  Recall if full retrocausality is allowed, then a local HV model cannot be distinguished from quantum theory \cite{lazarovici2015relativistic}. Interestingly, in Ref. \cite{Chaves2018} it is shown that quantum mechanics can be give bounds on types of retrocausal HV models that are allowed, by quantifying the degree of retrocausality contained in them. A measure of retrocausality $R$ is given by 
\begin{align}
R = \max \left[ {\frac{{{I_{DW}} - 3}}{4},0} \right],
\end{align}	
where ${I_{DW}}$ is the same dimensional witness used above. Using our experimental estimate we obtain $ R = 0.114 \pm 0003 $, in comparison to the ideal case where ${R^\text{Q}} = (\sqrt{2} - 1)/2 \approx 0.207$.  The meaning of this is that any retrocausal model with $ R \le 0.207 $ would not be able to reproduce the results that quantum theory is capable of producing.  Taking our estimate,  any retrocausal model with $ R \le 0.114 $ could not give consistent results to our experiment.


In summary, we have demonstrated a modified WDCE experiment and measured a device-independent witness to evaluate its consistency with a causally structured two-dimensional HV model.  We have found that the witness clearly shows that our experiment is inconsistent with such a HV model.  A key component in our experiment is to preserve the same causal structure as the model provided in Ref. \cite{Chaves2018}, which was achieved by separating Alice (preparer) and Bob (measurer) by space-like distances, which that there was no causal influence on each other.  We have also excluded HV theories which assume noise between Alice and Bob has been correlated in advance of the experiment, and we put a bound on the amount of retrocausality needed to explain our data without quantum mechanics.  Thus, the original WDCE experiment is salvaged, with no more reference to the rather heuristic notion of wave-particle duality, and the evidence for quantum theory against various classes of HV theory seems uncontestable.

This work was supported by the National Natural Science Foundation of China, the Chinese Academy of Sciences, and the National Fundamental Research Program. J. P. D was supported by the National Science Foundation of the United States.

H.-L. Huang and Y.-H. Luo contributed equally to this work.

$Note$ $added.$ During the final stages of manuscript preparation, we became aware of a similar work by Polino $et$ $al.$, which was carried out simultaneously and independently.

\bibliographystyle{apsrev4-1}
\bibliography{dc}

\end{document}